# Stress-Induced Variations in the Stiffness of Micro- and Nano-Cantilever Beams

R. B. Karabalin[1], L. G. Villanueva[1], M. H. Matheny[1], J. E. Sader[2], M. L. Roukes[1]

[1]*Kavli Nanoscience Institute, California Institute of Technology, Pasadena, California 91125,*

[2]*Department of Mathematics and Statistics, The University of Melbourne, Victoria 3010, Australia*

## ABSTRACT

The effect of surface stress on the stiffness of cantilever beams remains an outstanding problem in the physical sciences. While numerous experimental studies report significant stiffness change due to surface stress, theoretical predictions are unable to rigorously and quantitatively reconcile these observations. In this Letter, we present the first controlled measurements of stress-induced change in cantilever stiffness with commensurate theoretical quantification. Simultaneous measurements are also performed on equivalent clamped-clamped beams. All experimental results are quantitatively and accurately predicted using elasticity theory. We also present conclusive experimental evidence for invalidity of the longstanding and unphysical axial force model, which has been widely applied to interpret measurements using cantilever beams. Our findings will be of value in the development of micro- and nano-scale resonant mechanical sensors.

Physical or chemical adsorption onto the surface of micro- and nanoscale cantilevers is the basis for a rapidly growing field of biological sensing in the mechanical domain. Despite the remarkable achievements of the field [1, 2], it is surprising that understanding the mechanism by which biomolecules affect the physics of these mechanical structures remains elusive [3-17]. The adsorption process can result in a multiplicity of concurrent effects including: mass loading of the device [4, 5, 18], enhancement of surface elasticity [12, 13, 17, 19], increase in damping [20], and the imposition of surface stress [6-11, 15]. Numerous studies over the past 35 years have claimed that surface stress can cause significant variations in device stiffness [3-14, 17]. In contrast, other works ignore such effects, claiming that such stress loads do not affect the stiffness of cantilever devices, and argue for alternative interpretations. It is also striking that, within this extensive body of work [1, 2], controlled, quantitative measurements of the effects of surface stress on cantilever stiffness with commensurate theoretical interpretation have yet to be reported.

The original theoretical model of Lagowski *et al*. [15] suggested that a *net axial force* is induced along the beam axis upon application of stress – a load similar to the case of doubly-clamped beams derived within the framework of Euler-Bernoulli beam theory [21]. Lagowski *et al.* found that the *axial force model* was in quantitative and qualitative agreement with measurements. However, the axial force model as applied to cantilever beams has subsequently been shown to be in violation of Newton's 3[rd] law, i.e., it does not satisfy the fundamental physical principle of force equilibrium [16, 22, 23], contrary to doubly-clamped beams. In short, application of surface stress to a cantilever beam always induces stress of opposite sign within the beam material, resulting in zero net axial force along the beam. This phenomenon is contingent on the beam length greatly exceeding its width and thickness, a fundamental assumption of Euler-Bernoulli beam theory; see

Refs. [16, 22, 23]. Hence, Euler Bernoulli beam theory leads to the conclusion that cantilever beams should be insensitive to surface stress.

Though it was proven unphysical, numerous experimental studies have provided experimental data, seemingly consistent with the axial force model in micro- and nanoscale cantilever beams [6-9, 11, 15]. However, it is important to note that surface stress change in all such experimental studies is achieved using processes that may cause unspecified changes in the mechanical properties of the resonator [15, 24]. It is thus not entirely evident whether these previous measurements are due solely to stress changes at the cantilever beam surface, or effects of an unspecified nature. This situation contrasts strongly to clamped-clamped beams whose stress-induced stiffness change can be systematically observed and theoretically predicted [21, 25].

In this Letter we extend previous theoretical work, propose a new theoretical model and explain small but non-negligible changes in the resonant frequencies of cantilever beams due to application of stress. We also present the first systematic measurements of both cantilever and doubly-clamped beams using a robust and highly repeatable methodology, which shows remarkable agreement with theory. This resolves the above-mentioned long-standing debate and experimentally establishes the invalidity of the axial force model for cantilever beams.

In contrast to the case of doubly-clamped beams, the application of surface stress to a long and thin cantilever beam does not generate a net axial force because the longitudinal displacement is not constrained. However, a cantilever generates a net in-plane stress in the immediate vicinity of its supporting clamp [22], which can affect its resonance frequency. Expressions for the relative frequency shifts $\Delta f / f_R$ of doubly-clamped and cantilever beams due to the application of surface stress, and subsequent generation of net in-plane stress within the device, are shown in the second column of Table 1; termed the *stress effect*. We emphasize that all results in Table 1 are derived for thin beams.

| | $\Delta f / f_R$ (stress effect) | $\Delta f / f_R$ (geometric effect) |
|---|---|---|
| Doubly-Clamped Beam | $0.1475 \left(\dfrac{L}{h}\right)^2 \dfrac{(1-\nu)\sigma_s^T}{Eh}$ | $\dfrac{(1+2\nu)(1+\nu)}{2(1-\nu)} \dfrac{(1-\nu)\sigma_s^T}{Eh}$ |
| Cantilever Beam | $-0.042\nu \left(\dfrac{b}{L}\right)\left(\dfrac{b}{h}\right)^2 \dfrac{(1-\nu)\sigma_s^T}{Eh}$ | $\dfrac{1+2\nu}{1-\nu} \dfrac{(1-\nu)\sigma_s^T}{Eh}$ |

**Table 1** | Formulas determining the relative frequency shift of thin doubly-clamped beams (top row) and cantilever beams (bottom row) when a load $\sigma_s^T$ is applied. Both stress (left-column) and geometric effects are presented. While the latter effect is of equivalent magnitude for both boundary conditions, the former originates from: (i) a net axial force in the case of doubly-clamped beams, and (ii) from in-plane stress generation near the clamp for cantilevers. *L*, *b* and *h* are the length, width and thickness of the resonator respectively, *E* is the Young's modulus, *v* is the Poisson ratio, and $\sigma_s^T$ is an applied surface stress.

Application of a load to any elastic body changes its geometry due to elastic deformation. This *geometric effect* is typically ignored in the classical theory of linear elasticity. In the present context, application of surface stress induces a change in the beam length, width, thickness and density, which alter the resonant frequency of both clamped-clamped and cantilever beams. Expressions for the relative frequency shift due to this geometric effect are shown in the third column of Table 1. To account for the true geometry of the devices, finite element analysis is performed below. The complete effect is given by the sum of the stress and geometric effects.

The formulas in Table 1 indicate that the resonant frequencies of doubly-clamped beams are more sensitive to surface stress changes than cantilever beams (considering typical devices dimensions, $L > 10h$). For clamped-clamped beams, the stress effect listed in Table 1 dominates, whereas for cantilever beams the situation is more complex: the stress effect is dominant for thin structures ($h \ll b$), with geometric effects prevailing for thicker devices. Importantly, if the geometric effect dominates, the relative frequency shift $\Delta f / f_R$ does not change with length. Alternatively, if the stress effect prevails then $\Delta f / f_R$ strongly depends on length.

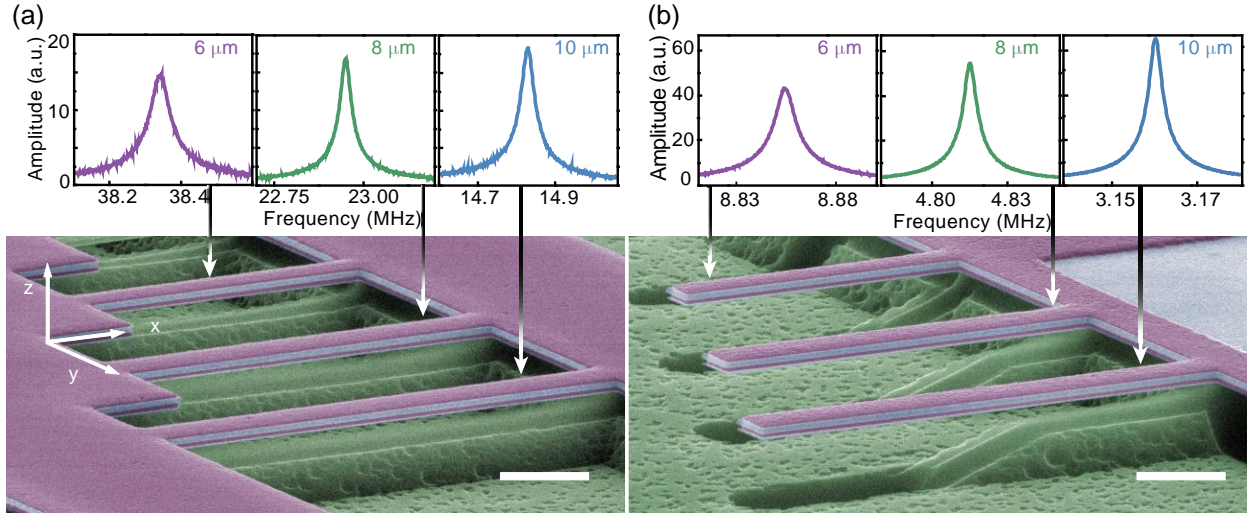

**Figure 1 | Resonant response of piezoelectric beams. a**, SEM micrograph of the doubly-clamped beams used for the experiments. On top of the micrograph, we show resonant responses of each of the beams, yielding resonant frequencies of 38.3 MHz (length 6 μm, purple), 22.9 MHz (8 μm, green) and 14.8 MHz (10 μm, blue). Experimental details are provided in the Supporting Information. **b**, SEM micrograph of the cantilever beams used for the experiments. Respective resonant responses are also shown for each cantilever, yielding natural frequencies of 8.85 MHz (length 6 μm), 4.82 MHz (8 μm), 3.16 MHz (10 μm). Both types of beams have the same composition (320 nm of total thickness) and width (900 nm). Lengths are 6, 8 or 10 μm for both types of devices, causing the boundary conditions to be the only difference, thus allowing proper comparison of the experimental results for the two configurations. Scale bars: 2 μm.

To examine the predictions of the above theoretical model, we perform systematic measurements on nanoscale cantilever and doubly-clamped beams of identical geometries. Our measurements are achieved by fabricating multilayered mechanical bars out of piezoelectric material, utilizing recent advances in nanoelectromechanical systems (NEMS) fabrication techniques [26]. Thus, stress changes in our devices are internally produced piezoelectrically, which allows for reproducible and controllable variation. It should be noted that in our experiment the stress is applied to a layer of finite thickness, rather than to the surface of the devices. To account for any differences that this may induce, we conduct rigorous numerical finite element simulations of both devices that includes their full structure and load (see Supplementary Information). Doubly-clamped beams with identical dimensions provide a benchmark for the measurements on cantilevers.

Cantilever and doubly-clamped nanoscale beams are fabricated on the same chip from a 320 nm 4-layer stack (comprised of: 20 nm aluminum nitride (AlN), 100 nm molybdenum (Mo), 100 nm AlN, and 100 nm Mo), using a process described elsewhere [26]. A standard wafer curvature measurement yields built-in stresses in all four layers close to zero. SEM micrographs of the devices are shown in Fig. 1. All beams possess an identical width $b = 900$ nm and total thickness $h_{tot} = 320$ nm, whereas their lengths $L$ are 6, 8 and 10 μm. The absence of curvature in cantilever beams indicates the absence of stress gradient in structural layer. Note that the only difference between cantilevers and doubly-clamped beams is that the boundary condition at one end is changed – material and geometric properties are otherwise identical. An electric field is produced in the active piezoelectric layer by applying a DC voltage between the top and bottom molybdenum layers. This induces an axial stress along the beam axis via its inherent piezoelectric properties. Ensuring the active piezoelectric layer is offset from the neutral axis of the beam enables the fundamental flexural mode of the device to be actuated through application of an additional AC voltage. Measurements of the resulting normal deflection are performed in vacuum using optical interferometry (see Supporting Information). Resonance measurements are shown in Fig. 1a and Fig. 1b for doubly-clamped beams and cantilevers respectively. We monitor the resonant frequency while changing DC voltage by means of a computer-controlled phase locked loop. Controllable variations in the stiffness of both types of devices due to stress changes are observed.

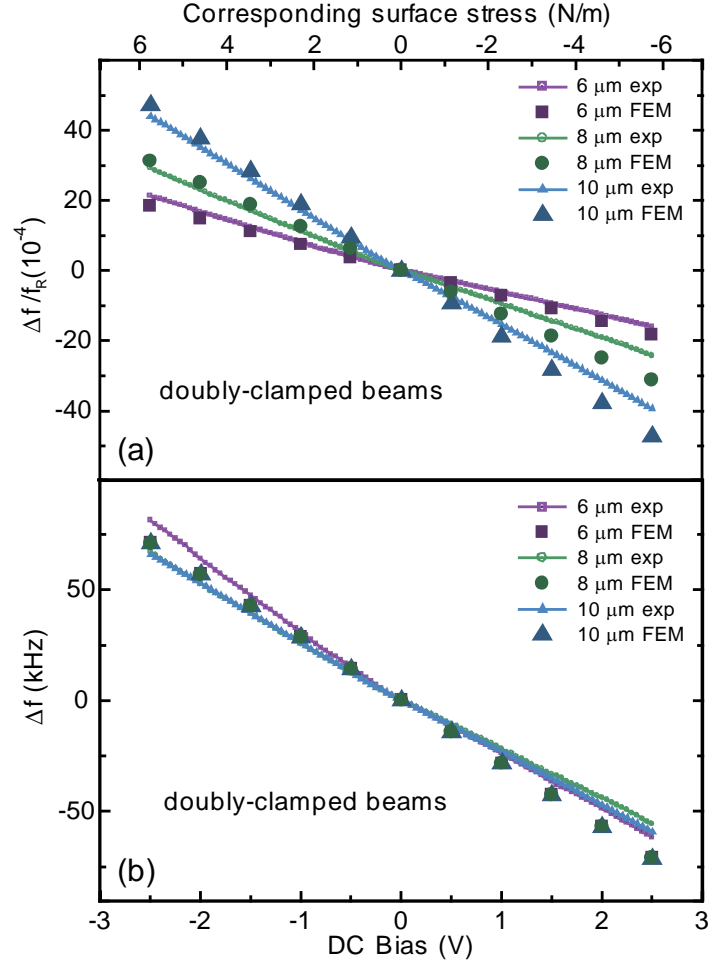

**Figure 2 | Frequency shift results for doubly-clamped beams. a**, Relative frequency shift $\Delta f/f_R$ for the three doubly-clamped beams in Fig. 1a, showing the dependence of $\Delta f/f_R$ on length (as predicted by theory). **b**, Absolute frequency shift $\Delta f$ (in kHz) for the same three beams. Experimental (lighter colors) and FEM results (scattered plots, darker colors) display excellent agreement. The measurements are shown as a function of the applied DC bias (in volts) and as a function of the corresponding surface stress, as calculated in the Supporting Information. The stress calculation requires estimation of the piezoelectric coefficient $d_{31}$; this was obtained by a linear fit of these experimental results, yielding a value of -2.5 pm/V.

We initially study the effect of an applied stress on the fundamental resonant frequencies of doubly-clamped beams. Formulas connecting resonance frequency shifts to surface stress are presented in Table 1 (see Supporting Information for corresponding expressions for piezoelectric loads). Measurement results of doubly-clamped beams are presented in Fig. 2, which clearly exhibit a linear variation in resonant frequency with voltage (stress) in the piezoelectric layer, as predicted theoretically. No variation of quality factor with bias voltage is observed. This provides strong validation for the robustness of the experimental methodology and the ability to tune the stiffness of doubly-clamped beams. Use of the resonant frequency shift expressions enables the axial force (or the corresponding surface stress, as commonly reported) to be calibrated against the applied voltage (see Supporting Information), which we determine to be 2.3 N/m per Volt. In agreement with theory (see Table 1), the relative frequency shift shown in Fig. 2a depends strongly on the beam length, scaling as $L^2$ ($f_R$ is the original resonant frequency), whereas the absolute shift in resonant frequency $\Delta f$ is found to be independent of the beam length (see Fig. 2b).

In Fig. 3 we present complementary results for the resonant frequency shift in cantilever beams. These nanomechanical devices have identical dimensions and material properties to the doubly-clamped beams in Fig. 2. Linear tuning of the resonant frequency with the applied piezoelectric layer voltage is also observed. However, these frequency shifts are two orders of magnitude smaller than those exhibited by doubly-clamped beams, even though the piezoelectric stress loads are identical – this constitutes direct experimental evidence of the invalidity of the unphysical axial force model for cantilever beams; see Eq. (S9). Not only the magnitude but also the scaling behavior of the resonant frequency shift $\Delta f$ with the beam length is inconsistent with the axial force model (see Supporting Information).

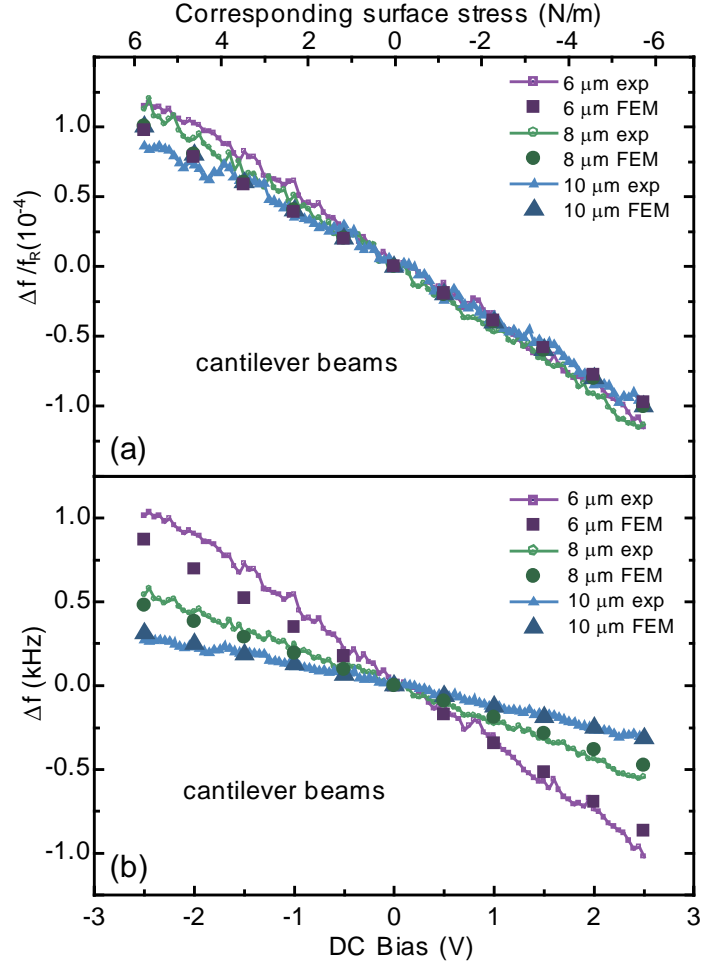

**Figure 3 | Frequency shift results for cantilever beams. a**, Relative frequency shift $\Delta f / f_R$ for the three cantilever beams in Fig. 1b, showing independence of length, which is consistent with the theoretical model. Both experimental (lighter colors) and FEM results (scattered plots, darker colors) are presented. FEM data are a result of combining both stress and geometric effects; these show excellent agreement with measurements. **b**, Absolute frequency shift $\Delta f$ (in kHz) for the same three beams. Comparison of these results with those in Fig. 2 demonstrates that the effect of stress on the resonant frequency is much smaller for cantilevers than doubly-clamped beams.

Notably, measurements of $\Delta f / f_R$ for cantilever beams display independence on the cantilever length and are identical in sign to those obtained for doubly-clamped beams; see Figs. 2a and 3a. This observation is completely consistent with the geometric effect listed in Table 1, which is expected to dominate since these devices are relatively thick ($b/h \sim 3$). Importantly, the formulas given in Table 1 are derived under the assumption of a thin beam ($h << b$), where the stress load is applied to an infinitesimal layer at the surface. To account for the true device geometry and properties, as noted above, we calculate the combined contribution from geometric and stress induced effects using full 3D finite element simulations (see Supplementary Information). The agreement between predictions from these simulations (shown as scatter plots in Figs. 2 and 3) and the experimentally measured data is within 15%. The apparent slight asymmetry between positive and negative voltages is within the experimental error (see Supplementary Information). These results constitute the first repeatable measurements and theoretical quantification of stress induced changes in the frequency shifts (stiffness) of cantilever beams. They also provide compelling experimental evidence for the invalidity of the axial force model, which has been widely applied to the interpretation of measurements made using cantilever beams.

It remains to understand why several previously reported cantilever measurements display results that are inconsistent with the present findings [6-9, 11, 14, 15]. These studies report stress-induced stiffness changes that appear to be in agreement with the unphysical axial force model that erroneously predicts effects much larger in magnitude than the results reported here. These previous, now anomalous, measurements were typically performed using surface adsorption or modification processes. While these processes are known to induce surface stress, additional unspecified and uncontrolled effects are also possible. There is certainly the distinct possibility of over-layer formation, since material is adsorbed to [12, 13, 17, 19] or removed from the beam surface [15]. Some studies have shown excellent agreement between classical composite beam theory and such measurements [12, 13, 17, 19], while other measurements remain unexplained [6-9, 14, 15]. Our findings strongly suggest that previous reports of stress-induced changes in cantilever beams originate from other uncontrolled surface phenomena.

Miniaturization of NEMS technologies is key to enhanced sensitivity and ultrafast measurements. Our study establishes that as cantilever thickness is reduced, a transduction mechanism arises that is different to the one observed in the present devices. Specifically, we predict that cantilever devices made of ultrathin materials (such as graphene) allow for gigantic tunability in their resonance properties. Such materials would exhibit an additional inverse-squared thickness dependence, enabling very strong modification of their stiffness. This could ultimately permit the development of UHF bandpass cantilever filters, with broad control of their frequency range for applications in biosensing, telecommunications and medical technologies.

We would like to thank X. L. Feng, J. Xiang, M. J. Lachut for useful suggestions and discussions. We would also like to thank E. Defaÿ, G. Le Rhun, C. Marcoux from CEA-LETI for providing us with Aluminum Nitride material. L.G.V. acknowledges financial support from the European Commission (PIOF-GA-2008-220682) and Prof. A. Boisen. J.E.S. acknowledges support from the Australian Research Council grants scheme.